\documentclass[runningheads]{llncs}
\usepackage{graphicx}
\usepackage{bibnames}
\usepackage{multirow}
\usepackage{tabularx}
    \newcolumntype{L}{>{\raggedright\arraybackslash}X}
\begin{document}
\title{Radiogenomics of Glioblastoma: Identification of Radiomics associated with Molecular Subtypes}
%
%

\author{Navodini Wijethilake\inst{1} \and
Mobarakol Islam\inst{2} \and
Dulani Meedeniya\inst{1} \and 
Charith Chitraranjan\inst{1} \and
Indika Perera\inst{1} \and
Hongliang Ren\inst{2}
}
\authorrunning{N. Wijethilake et al}
\institute{
Dept. of Computer Science Engineering, University of Moratuwa, Srilanka\\
\and
Dept. of Biomedical Engineering, National University of Singapore, Singapore\\
}

%

\maketitle              
\begin{abstract}
Glioblastoma is the most malignant type of central nervous system tumor with GBM subtypes cleaved based on molecular level gene alterations. These alterations are also happened to affect the histology. Thus, it can cause visible changes in images, such as enhancement and edema development. In this study, we extract intensity, volume, and texture features from the tumor subregions to identify the correlations with gene expression features and overall survival. Consequently, we utilize the radiomics to find associations with the subtypes of glioblastoma. Accordingly, the fractal dimensions of the whole tumor, tumor core, and necrosis regions show a significant difference between the Proneural, Classical and Mesenchymal subtypes. Additionally, the subtypes of GBM are predicted with an average accuracy of 79\% utilizing radiomics and accuracy over 90\% utilizing gene expression profiles. 

\end{abstract}
\section{Introduction}

Glioblastoma (GBM) is a common malignant tumor that occurs mostly in adults, $>$ 50 years of age and accounts for 14.7\% of primary brain and central nervous system tumors. GBM has the highest incidence rate of 3.21 per 100000 population, concerning other malignant tumors. Despite the treatment followed by the diagnosis, GBM patients have poor survival, with a median survival of one year and a five-year survival rate of 5\% according to a study done from 2010 to 2015 \cite{ostrom2018cbtrus}. This aggressive behavior of GBM is mainly due to the molecular level heterogeneity and complexity. Thus, identifying the prognostic genetic biomarkers is essential to decide the appropriate therapies and the treatment required for each GBM patient.

Accordingly, many research groups have focused on identifying genetic biomarkers through gene expression profiling, copy number alterations, mutation studies. Thus, gene mutations such as Tumor Protein p53 (TP53), Retinoblastoma protein (RB1), Neurofibromin (NF1), and Telomerase Reverse Transcriptase (TERT) have shown associations with GBM \cite{verhaak2010integrated}. Moreover, Isocitrate dehydrogenase 1 (IDH1) mutations are common in secondary GBM, developing from a low-grade glioma. Other than that, overexpression of Epidermal growth factor receptor (EGFR), Phosphatase and Tensin homolog (PTEN), Platelet-derived growth factor receptors (PDGFR) genes found to have had an impact on GBM prognosis \cite{delgado2016survival}. Consequently, Verhaak et al. \cite{verhaak2010integrated} have identified 4 GBM subtypes, named Proneural, Neural, Mesenchymal, and Classical, based on gene expression profiles and other genetic alterations. 

A classical subtype can be recognized with the overexpression of EGFR and the loss of chromosome 10. In contrast, Mesenchymal subtype is mostly related to the expression of the NF1 gene and PTEN deletions. The patterns in these two subtypes are commonly seen in astrocytes. The proneural subtype is recognizable with the alterations of PDGFR, IDH1, and TP53 in oligodendrocytes, whereas the expressions in neurons characterize the Neural subtype. Nevertheless, the Proneural subtype has a better prognosis for the other subtypes \cite{chen2018immune,delgado2016survival}. The Neural subtype existence has been revisited and removed as it has emerged through contamination with the non-tumor cells \cite{sidaway2017glioblastoma,wang2017tumor}. The specific treatments are given for each subtype by the clinicians, and therefore, it is important to identify the subtypes at first. However, due to the costs and the invasive approach, gene expression profiling analysis is not routinely performed on GBM patients and is an existing challenge. 

Alternatively, Magnetic Resonance Imaging (MRI) is a widely used noninvasive imaging tool to identify gliomas. With different types of MRI modalities, such as T1, T1 Contrast, T2, and Flair (T2 fluid-attenuated inversion recovery), underlying tissues of gliomas can be visualized for diagnosis \cite{delgado2016survival}. With the advancements in the field of artificial intelligence, deep learning-based tools have been developed for segmenting the gliomas using MRI. This further has extended the area of radiomics through extracting features for diagnosing, predicting survival, etc \cite{islam2018glioma}. Moreover, the field of radiogenomics was emerged by analyzing the associations between genetic biomarkers and radiomics.
Nonetheless, to address the current limitations in identifying the subtypes, imaging biomarkers can be in-cooperated. As we mentioned above, there are limitations exploiting gene expression profiling routinely, due to invasive behavior and the cost. Therefore, this motivated our study to predict the subtype, with noninvasive biomarkers obtained from MRI imaging, commonly used clinical practice. 

In several studies related to various cancer types, imaging features are utilized to determine the underlying molecular subtype \cite{mazurowski2014radiogenomic,sutton2016breast}. Nonetheless, different statistical analyses are performed to identify imaging biomarkers related to each subtype of GBM patients separately \cite{naeini2013identifying}. Prior studies have also focused on predicting the subtypes as a 4 class classification \cite{macyszyn2015imaging}. The novelty of our research is, despite imaging biomarkers related to a single subtype, we identify imaging biomarkers that are different between the three main GBM subtypes. Additionally, we classify existing subtypes, Proneural, Mesenchymal, and Classical, separately as a binary classification task with accuracy over 80\%, since the Neural subtype does not exist anymore. 

In this work, we use the TCGA-GBM dataset with both MRI and gene expression data for analyzing the associations with the survival of each feature. The main contribution of our study is to identify radiomic features that are significantly different between the subtypes of GBM patients. Moreover, we utilize those features to predict the existing subtypes, Proneural, Mesenchymal, and Classical separately. We perform cross-validation to verify our results.

\section{Methods}
\subsection{Dataset}

The radiogenomic analysis is performed on The Cancer Genome Atlas (TCGA) \footnote[1]{https://www.cancer.gov/tcga} GBM dataset. We have obtained 202 cases used in the study \cite{verhaak2010integrated} with the subtype of each instance, including the gene expression profiles. Out of 202, 59 cases have the corresponding MRI data, with T1 contrast and flair modalities. MRI images are skull stripped initially with Matlab. Then, registered to 155 x 155 x 240 3D image volume with the origin at (0,-239,0), making the metadata similar to BraTS 2019 dataset, with open-source 3D slicer software \cite{pieper20043d}.

The gene expression profiles of 59 patients consist of 1740 genes for each patient. Moreover, for these patients, subtypes class (Classical, Mesenchymal, Proneural, or Neural) and the overall survival from the diagnosis are also filtered. 

For the fully automated segmentation with Deep Learning, Brats 2019 dataset is used for training the segmentation model using T1 Contrast and Flair modalities. The segmentation includes three sub-regions, necrosis, enhancing tumor region and edema region, annotated manually in the BraTS dataset. 

\subsection{Segmentation}


We utilize the 3D UNet architecture \cite{3Dunet2020Islam}, with 3D attention module \cite{roy2018concurrent}, that we proposed for BraTS challenge 2019. The 3D spatial and channel squeeze and excitation modules learn more important and meaningful feature maps through spatial and channel re-calibration. In this module, the channel wise dependencies are captured through 1x1xC convolution of the feature map (with the dimensions of HxWxC), and the spatial feature correlation are captured through max-pooling the feature map. This boosts the quality of segmentation prediction, minimally changing the model complexity.  

155 x 155 x 240 3D MRI volume is given as the input of the segmentation model, to obtain the segmentation feature map. The model is trained with the BraTS manually annotated feature maps. The segmentation of the TCGA dataset is achieved with the trained model giving the FLAIR and T1 Contrast modalities as input. 

\subsection{Feature Extraction}
The segmentation of the GBM is followed by the imaging feature extraction from the tumor's relevant sub-regions. Intensity, Geometric and volume features are extracted from the segmented sub-regions of the tumor. 

The intensity and texture-based features, kurtosis, the histogram of oriented gradients, entropy, mean intensity, are extracted for the necrosis, enhancement, tumor core (necrosis + enhancement region), whole tumor (tumor core + edema) regions. These histogram features are extracted from both T1 contrast and Flair images, since both of them, magnify different sub-regions of the tumor. For example, T1 contrast provides a clear picture of the enhancement region. 

The extracted geometric features include the major axis length, minor axis length, centroid coordinates, Inertia tensor, Eigenvalues of the inertia tensor. Moreover, the volume features are extracted with the bounding box and with the factual dimensions. These features are extracted for the necrosis, tumor core, and the whole tumor regions. Totally 122 imaging features are extracted from the MRI images.

\subsection{Statistical Analysis}
Our initial experiments are to find the associations between genomics and radiomics. Thus, we have utilized the Pearson's correlation to find the association among imaging features and the gene expression levels of GBM patients. Further, we use both feature types, radiomics and genomics, separately to find the correlation with the overall survival of glioma patients. For that also, Pearson's correlation is used. Nevertheless, univariate and multivariate cox proportional hazard analysis is performed to identify the radiogenomic features associated with survival.

Next, we have focused on identifying the imaging biomarkers associated with the molecular subtypes of GBM. First, to clarify the normality of the radiomic features, the Kolmogorov-Smirnov test \cite{naeini2013identifying} is utilized, and the features with $p < 0.05$ are identified as features with a normal distribution. Accordingly, we have performed Kruskal Wallis non-parametric test \cite{naeini2013identifying} to identify the significantly different imaging biomarkers between the four subtypes, with the features that are not normally distributed. Additionally, to identify the subtype pairs that are significantly different Wilcoxon test \cite{follia2019integrative} is applied between each pair. The Bonferroni correction is performed on the multiple comparison tests to avoid the familywise error rate. For visualization of the significance and the feature distribution, GraphPad prism software is used.

\subsection{Subtype Predictive model}
We have used the significantly different features between subtypes to predict the subtype using Machine Learning algorithms. All the features are standard normalized before training the learning algorithm. Linear Kernel function of Support Vector Machine (linear-SVM), radial basis kernel function of Support Vector Machine (r-SVM), polynomial kernel function of Support Vector Machine (p-SVM), Random Forest Classification (RFC), Decision Tree (DT) and Logistic Regression (LR) are utilized for predicting Classical, Mesenchymal and Proneural subtypes separately with the imaging biomarkers. The parameters used in the above learning models are given in the Table \ref{tab:parameters}. This is performed as a 2 class classification (subtype/non-subtype). 

Nevertheless, linear-SVM, r-SVM, p-SVM, DT and RFC are used to predict the subtypes with the differentially expressed genes between the subtypes. This is performed as a 4 class classification (Mesenchymal, Proneural, Neural, and Classical). 

Further, four-fold cross-validation is performed to validate the performance of all the subtype predictions. All the experiments are performed with e1071 R package \cite{svmpackage} and WEKA software \cite{weka}.

\begin{table}[!h] \centering
\caption{Parameters of the subtype prediction learning models}\label{tab:parameters}
\begin{tabular}{|l|l|}\hline
ML tool &   Parameters\\\hline
l-SVM & linear kernel function \\\hline
r-SVM & radial basis function \\\hline
p-SVM & polynomial kernel function \\\hline
DT &  depth of the tree: 5 \& Number of leaves: \\\hline
RFC & number of trees: 100  \\\hline
\end{tabular}
\end{table}
\section{Results}
\subsection{Correlation between radiomics, genomics and overall survival}

Out of the 119 radiomic features extracted, two radiomics showed a significant low positive correlation ($p<0.05$ \& correlation $>$ 0.3) with the overall survival in days. These two features are the kurtosis of the enhancement region extracted from T1 contrast and FLAIR MRI modalities. These two features also gave statistically significant ($p<0.05$) negative coefficients for the Cox proportional hazard analysis. Thus, the large kurtosis of the enhancement is associated with longer survival.

The correlation between the gene expression level of 1023 genes and the overall survival has also been analyzed. 25 genes out of 1023 genes, were significantly correlated with the overall survival. Three of these genes, Endothelin Receptor Type A (EDNRA), Olfactomedin-like protein 3 (OLFML3), and Collagen Type III Alpha 1 Chain (COL3A1), gave a low positive correlation. The rest of the genes showed a low negative correlation. Those genes are given in the Table \ref{tab:neg_corr_genes}.

\begin{table}[!h] \centering
\caption{The genes negatively associated with the overall survival obtained from the correlation study.}\label{tab:neg_corr_genes}
\begin{tabularx}{\linewidth}{|L|}
\hline
Low negatively correlated genes\\ \hline
Pirin (PIR)\\ 
Secretogranin-2 (SCG2)  \\ 
Insulin Like Growth Factor Binding Protein 3 (IGFBP3) \\ 
Mucosa-associated lymphoid tissue lymphoma translocation protein 1 (MALT1)\\ 
Tubulin alpha-4A (TUBA4A)\\ 
Growth Differentiation Factor 15 (GDF15)\\ 
Tetraspanin-13 (TSPAN13)\\ 
Neural precursor cell expressed developmentally downregulated gene 4-like (NEDD4L)\\ 
Endoplasmic Reticulum Aminopeptidase 2 (LRAP) \\ 
Phospholipid Scramblase 1 (PLSCR1) \\ 
Ankyrin Repeat And MYND Domain Containing 2 (ANKMY2)\\
IQ Motif Containing G (IQCG)\\ 
Rho Family GTPase 3 (RND3)\\ 
Radical S-Adenosyl Methionine Domain Containing 2 (RSAD2)\\  
Pyruvate Dehydrogenase Kinase 1 (PDK1)\\ 
Regucalcin (RGN)\\ 
Dihydrouridine Synthase 4 Like (DUS4L)\\  
Oligoadenylate Synthetase 1 (OAS1)\\ 
FAM3 Metabolism Regulating Signaling Molecule C (FAM3C)\\  
Membrane Palmitoylated Protein 6 (MPP6)\\
\hline
\end{tabularx}
\end{table}

Nevertheless, these features were recognized with a Hazard ratio of $>$ 1 and $p<0.05$ with the univariate cox-PH analysis. 

Besides, the correlations between the imaging and genomic markers are also assessed. Thus, the expression level of the TIMP Metallopeptidase Inhibitor 4 (TIMP4) gene showed a moderate positive correlation (correlation $>$ 0.5) with the fractal dimension of the necrosis region. Similarly, the necrosis region's extent had a moderate positive relationship with the acyl-CoA oxidase 2 (ACOX2) gene expression level. 

\subsection{Imaging biomarkers associated with molecular subtypes}

Initially, the Kolmogorov-Smirnov test was applied to obtain the normality of the radiomics features. 38 features out of 119 radiomic features, were significantly normally distributed. Thus, for the remaining non-normal features, the Kruskal Wallis test was applied. Therefore, 12 features that were significantly ($p<0.05$) different between 4 subtypes are identified. These features include the fractal dimensions of the necrosis, whole tumor, and tumor core regions. All these features are given in Table \ref{tab:KW-radio}, with the test statistic and p-value, obtained. 

\begin{table}[!h] \centering
\caption{Kruskal Wallis statistics for radiomic features ($p<0.05$) between 4 subtypes.}\label{tab:KW-radio}
\begin{tabular}{|l|l|l|}
\hline
Feature &  test statistic & p value\\
\hline
tumor core bounding box 1 & 8.8889 & 0.0308 \\
tumor core fractal dimension 3 & 7.9312 & 0.0474 \\
whole tumor fractal dimension 4 & 7.8459  & 0.0493 \\
whole tumor fractal dimension 5 & 7.9820 & 0.0463 \\
kurtosis of necrosis  - Flair & 9.197 & 0.0267 \\
bounding box - 1 of necrosis  & 9.263 & 0.0259 \\
bounding box - 3 necrosis & 9.0134 & 0.0291 \\
fractal dimension 1 of necrosis & 8.887 & 0.0308 \\
fractal dimension 2 of necrosis & 9.1918 & 0.0268 \\
fractal dimension 3 of necrosis &10.418 & 0.0153 \\
fractal dimension 4 of necrosis & 9.1716 & 0.027 \\
fractal dimension 5 of necrosis & 8.937 & 0.0301  \\ 
\hline

\end{tabular}
\end{table}

\begin{figure}[!h]
\includegraphics[width=\textwidth]{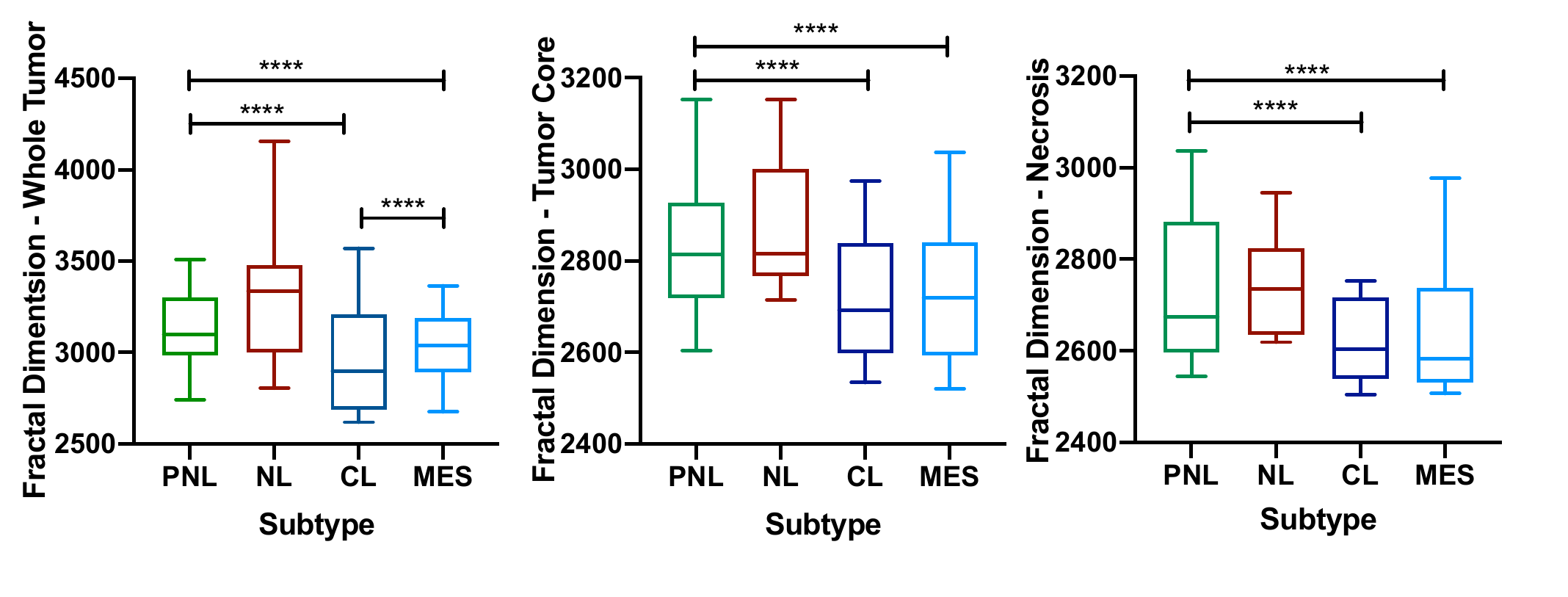}
\caption{The comparison of the fractal dimensions of whole tumor, tumor core and necrosis regions between 4 subtypes of Glioblastoma.} \label{fig1:MWU}
\end{figure}

This suggests that these features were significantly different between some subtypes. To identify those differences between subtypes, the Wilcoxon test was utilized, pairwise. The fractal dimension of the necrosis, tumor core, and whole tumor region showed a significant difference ($p < 0.0001$) between Proneural and Classical subtypes and also a considerable difference ($p<0.0001$) between Proneural and Mesenchymal subtypes. Moreover, the fractal dimension of the whole tumor showed a substantial difference between Classical and Mesenchymal subtypes. The associations and the distributions of the fractal dimensions are shown in the Fig \ref{fig1:MWU}.

\subsection{Subtype Prediction}
The significantly different 12 radiomic features identified through the Kruskal-Wallis test was used for this subtype prediction. First, the Mesenchymal subtype (Mesenchymal - 23 cases, Non-Mesenchymal - 36) was predicted with ML tools, and SVM outperformed the other algorithms with an accuracy of 62.7119\%, along with  61\% precision and 62.7\% recall. Next, the Classical subtype (Classical - 12, Non-Classical - 47) is predicted, and the SVM outperformed the other three algorithms with fourfold validation accuracy of 85.3\%. Nevertheless, the Prediction of Proneural subtype (Proneural - 16, Non-Proneural - 43) was performed to obtain a four-fold accuracy of 81.82\% accuracy with SVM. The performance comparison of the binary classification of each subtype are given in Table \ref{tab:summary_results}.

\begin{table}[!h] \centering
\caption{Prediction of subtypes as a binary classification using Radiomics. Acc, Prec, Rec are denotes for accuracy, precision and recall.}\label{tab:summary_results}
\begin{tabular}{|p{12mm}|l|c|c|c|c|c|c|c|c|c|c|c|c|c|c|}
\hline
\multirow{2}{*}{ML tool} & \multicolumn{4}{c|}{Classical}& \multicolumn{4}{c|}{Proneural}& \multicolumn{4}{c|}{Mesenchymal} \\ \cline{2-13}
 &  Acc & Prec & Rec& F1& Acc & Prec & Rec&  F1&Acc & Prec & Rec &F1\\
\hline
l-SVM & 85.3\% & 91.4\%& 88.8\% &84.5\%  & 81.8\% & 88.8\%&  88.8\% &86.3\%& 82.0\%  & 79.2\%&  82.0\% &80.2\%\\ 
r-SVM & 82.3\% & 84.1\%& 81.2\%  & 81.2\% & 83.1\%&  83.6\% & 85.3\% & 84.6\%&  81.4\% &83.2\%&80.2\%&78.3\%\\ 
p-SVM & 78.8\% & 80.2\%& 79.4\%  & 79.4\% & 81.5\%&  83.2\% & 79.4\% & 81.4\%&  82.4\% &79.4\%&77.5\%&78.3\%\\ 
DT & 83.1\%& 83.1\%&  83.1\% &81.9\%& 66.1\%& 63.3\%&  66.1\% &64.7\%& 57.6\% & 58.7\%&  57.6\%&58.7\% \\
RFC & 79.7\% & 77.3\%&  79.7\% &78.3\%& 69.4\% & 62.7\%& 69.5\%&68.6\%& 61.0\%& 60.2\%&  61.0\% &61.4\% \\ 
LR &  77.9\%& 80.1\%& 78.0\%&76.4\%& 64.4\% & 65.1\%& 64.4\% &65.1\%& 57.6\% & 56.7\%&  57.6\% &55.4\%\\
\hline
\end{tabular}
\end{table}

The subtype prediction with genomics was executed as a generalized method for predicting any of the 4 subtypes. However, since the Neural subtype no longer exists, this model might require slight changes in current exploration. Thus, the SVM gave the highest accuracy of 94.91\% with four-fold cross-validation. The RFC also performed comparatively better than the DT algorithm with an accuracy of 89.93\%, verifying the usage of gene expression despite the acquisition.

\begin{table}[!h] \centering
\caption{Prediction of Molecular subtypes as a 4 class classification with Genomics}\label{tab:Proneural}
\begin{tabular}{|l|c|c|c|c|c|}
\hline
ML method &  Accuracy & Precision & Recall & F1-score\\
\hline
l-SVM & 94.91\% & 95.40\%& 94.90\%& 92.67\% \\ 
r-SVM & 92.47\% & 93.29\%& 91.50\% & 91.43\%\\ 
p-SVM & 89.65\% & 90.41\%& 88.60\% & 89.65\%\\ 
DT & 64.40\%& 63.30\%&  64.40\% & 63.30\% \\
RFC & 89.83\% & 90.81\%&  89.80\% & 88.70\% \\ 
\hline
\end{tabular}
\end{table}

\section{Discussion}
In our retrospective study, we analyze the associations between radiomics, genomics, and overall survival of GBM patients. Our results showed that the kurtosis of the enhancement region has a positive correlation with survival. Kurtosis represents to what extent the intensity of this particular region deviates from a perfect normal Gaussian distribution. Nevertheless, the enhancement region has shown relationships with the survival, in previously done GBM related studies \cite{carrillo2012relationship,gutman2013mr}. 

While analyzing genes associated with survival, we identified that the COL3A1 gene expression level correlates with overall survival. COL3A1 gene is considered a prognostic marker of GBM patients \cite{gao2016col3a1} and have shown associations with GBM prognosis. Further, the TIMP4 gene showed a correlation with the fractal dimension of the necrosis region. According to previous studies, \cite{rorive2010timp}, TIMP4 overexpression cause shorter survival in GBM patients. Thus, we can hypothesize that this overexpression in GBM patients can be approximated through the necrosis region's fractal dimension. Another major finding of our study is the differentiation of fractal dimension between 4 subtypes. The fractal dimension provides an assessment of each region's complexity, where we used a box-counting approach to obtain this dimension. 

The radiomic features that predict survival provide a suitable platform to predict the subtypes according to our results. Despite the technological advancements, obtaining a consistent gene expression profiles is hard due to the technical shortcomings. Therefore, identifying the subtype with radiomics helps clinicians determine the subtype precisely, with the acquisition of MRI that has been used in clinical practice commonly, followed by a fully automated segmentation and feature extraction. 

Originally, GBM is recognized with four subtypes by Verhaak et al. \cite{verhaak2010integrated}. Recently, Wang et al. \cite{wang2017tumor} define that only Classical, Mesenchymal, and Proneural subtypes exist, and the previously described Neural subtype was identified through contamination with nontumor cells. Therefore, we performed subtype prediction as a binary classification of the Classical, Mesenchymal, and Proneural subtypes separately. Another limitation of our study is the sparse dataset. Yet, we utilized the same dataset to analyze the overall survival of GBM patients with an outperforming accuracy \cite{wijethilake2020radiogenomics}. However, we believe an extended dataset, with both MRI and Gene expression profiling, will lead to a robust subtype predictive model. 

\section{Conclusion}
In this study, we address a solution for subtype prediction without employing high dimensional genomics, which is inconsistent with the platform acquired. We identified distinctive radiomics between subtypes, that reflects the underlying molecular level gene alterations between the subtypes. Accordingly, we developed a predictive subtype model, with machine learning for GBM patients. 
In conclusion, our work delivers a promising subtype predictive model that will lead to better clinical maintenance of GBM patients. 

\bibliographystyle{splncs04}
\bibliography{mybib}

\clearpage
\end{document}